\documentclass[twocolumn]{aastex631}
\usepackage{graphicx}
\usepackage{amsmath}
\usepackage{color}
\usepackage{comment}
\usepackage{units}
\usepackage[normalem]{ulem}
\usepackage{acronym}
\usepackage{braket}

\newcommand{\SPA}{School of Physics and Astronomy, Monash University, Clayton VIC 3800, Australia}
\newcommand{\OzGravMonash}{OzGrav: The ARC Centre of Excellence for Gravitational Wave Discovery, Clayton VIC 3800, Australia}
\newcommand{\CanInst}{Canadian Institute for Theoretical Astrophysics, 60 St George St, University of Toronto, Toronto, ON M5S 3H8, Canada}
\newcommand{\DavDun}{David A. Dunlap Department of Astronomy and Astrophysics, University of Toronto, 50 St. George St., Toronto ON. M5S 3H4, Canada}
\newcommand{\DepPhysTor}{Department of Physics, University of Toronto, 60 St. George St., Toronto, ON M5S 3H8, Canada}
\newcommand{\MaxPlanck}{Max-Planck-Institut für Gravitationsphysik, Albert-Einstein-Institut, Callinstr. 38, D-30167 Hannover, Germany}
\newcommand{\LeibUni}{Leibniz Universität Hannover, D-30167 Hannover, Germany}

\begin{document}
\title{Is GW231123 a hierarchical merger?}
\author{Lachlan Passenger}
\affiliation{\SPA}
\affiliation{\OzGravMonash}
\author{Sharan Banagiri}
\affiliation{\SPA}
\affiliation{\OzGravMonash}
\author{Eric Thrane}
\affiliation{\SPA}
\affiliation{\OzGravMonash}
\author{Paul D. Lasky}
\affiliation{\SPA}
\affiliation{\OzGravMonash}
\author{Angela Borchers}
\affiliation{\MaxPlanck}
\affiliation{\LeibUni}
\author{Maya Fishbach}
\affiliation{\CanInst}
\affiliation{\DavDun}
\affiliation{\DepPhysTor}
\author{Claire S. Ye}
\affiliation{\CanInst}

\acrodef{GW}[GW]{gravitational-wave}
\acrodef{BBH}[BBH]{binary black hole}

\begin{abstract}
    The binary black hole merger GW231123 is both the most massive gravitational-wave event observed and has the highest component spins measured to date. The dimensionless spins of the more massive (primary) and less massive (secondary) black holes are measured to be $\chi_1 = 0.90^{+0.10}_{-0.19}$ and $\chi_2 = 0.80^{+0.20}_{-0.51}$ ($90\%$ credible intervals), respectively. Its large mass and extremal spins are challenging to explain through standard binary stellar physics, though a flurry of hypothetical scenarios have been proposed. Hierarchical assembly---i.e., mergers of black holes that are themselves formed from previous generations of mergers---is generally a promising way to explain massive and rapidly spinning black holes. Here, we investigate the possibility that GW231123 was assembled hierarchically in a dense star cluster as the merger of two second-generation black holes. Taking the inferred spin values at face value, we find that it is possible ($p\approx 5\%$) that a compact binary with component spins like GW231123 could form in a cluster from hierarchical assembly. 
\end{abstract} 
\section{Introduction}
Of the $\sim{200}$ \ac{GW} events \citep[][]{collaboration_gwtc-40_2025} detected so far by the LIGO \citep[][]{theligoscientificcollaborationAdvancedLIGO2015}, Virgo \citep[][]{acerneseAdvancedVirgo2nd2015a}  and KAGRA \citep[][]{akutsuKAGRAGenerationInterferometric2019} collaboration (LVK), GW231123 \citep[][]{collaboration_gw231123_2025} stands out as a potential challenge to our understanding of binary formation. 
It is the most massive event observed to date, with component masses $m_1=137^{+22}_{-17}$ $\text{M}_\odot$ and $m_2=103^{+20}_{-52}$ $\text{M}_\odot$ (all uncertainties correspond to 90\% credibility).\footnote{For an alternative point of view, see \cite{mandel_what_2025} who considers the possibility that the masses associated with GW231123 could be biased by a tendency to look at the most extreme events.} 
Its dimensionless spins are consistent with the maximum allowed spin in general relativity: $\chi_1 = 0.90^{+0.10}_{-0.19}$ and $\chi_2 = 0.80^{+0.20}_{-0.51}$.

The formation of high-mass black holes is expected to be difficult through stellar collapse due to pair-instability and pulsational pair-instability processes in massive stars \citep[e.g.][]{hegerNucleosyntheticSignaturePopulation2002, woosleyPulsationalPairInstability2007, 
belczynskiEffectPairinstabilityMass2016b, woosleyPulsationalPairinstabilitySupernovae2017,woosleyEvolutionMassiveHelium2019, woosley_birth_2020, powell_final_2021, woosleyPairinstabilityMassGap2021, chen_multidimensional_2023, sykes_long-time_2024}, leading to a putative black-hole mass gap of around $50-130 M_\odot$. The precise bounds of this gap are not clearly known due to uncertainties in nuclear reaction rates \citep[e.g.,][]{farmerMindGapLocation2019, costaFormationGW190521Stellar2021} and further depend on details of mass fallback and envelope retention during core collapse \citep[e.g.,][]{winchPredictingHeaviestBlack2024}. 
If GW231123 formed from isolated stellar binary evolution, it is possible that its component black holes have masses \textit{above} the mass gap, where massive stars, stabilized by photon disintegration processes, can directly collapse to black holes.

In models with efficient angular-momentum transport in massive stars, isolated stellar-mass black holes are expected to be born with very low spins ($\chi_\text{birth} < 0.01$)
\citep[e.g.,][]{qin_spin_2018, fuller_most_2019}. \citet[][]{fuller_most_2019} achieve this using a Tayler-instability transport prescription motivated by asteroseismology in low-mass stars, though the extent to which this can be extrapolated to massive stars is uncertain.

However, analyses of the population of binary black holes are increasingly difficult to reconcile with the idea that many/most black holes have negligible spin \citep{callister_no_2022, mould_which_2022, gwtc-3_hui,single-spin,both_spin,gwtc-4_pop, Banagiri:2025dxo, szemraj_disentangling_2025}.
This therefore necessitates astrophysical pathways through which black holes either get spun up or retain a significant fraction of the angular momentum of their stellar progenitors. 

One formation scenario that can produce rapidly-spinning black holes is the hierarchical-merger channel \citep[for a review, see][]{gerosa_hierarchical_2021}. In dense environments, the merger product of a binary black hole coalescence can be retained and be subject to further mergers through dynamical capture. As this remnant inherits a significant fraction of the orbital angular momentum of its progenitors \citep[e.g.,][]{buonanno_estimating_2008}, only a narrow range of values are possible for its spin.
For comparable-mass black holes with moderate spins, the remnant spin is typically $\chi_f \approx 0.7$ \citep[][]{tichy_final_2008}. 
Thus, large black hole spins of $\chi\approx 0.7$ measured in merger progenitors can be a signature of hierarchical assembly. 

That said, the signature is far from unique: gas accretion in close binaries \citep[e.g.,][]{belczynski_most_2020} from the accretion disk of an active galactic nucleus \citep[e.g.,][]{tagawa_spin_2020,  mckernan_ligovirgo_2022, vajpeyi_measuring_2022} or in a dense star cluster \citep[][]{kiroglu_beyond_2025}, chemically homogeneous evolution \citep[e.g.,][]{marchant_new_2016, marchant_upper_2024, stegmann_resolving_2025} or tidal spin up \citep[e.g.,][]{bavera_approximations_2021, bavera_probing_2022, ma_tidal_2023, qin_merging_2023} may also lead to black-hole components with spins as high as $\chi = 0.7$. However, these formation channels should also produce black holes with a wide range of spins. 

Various hypotheses have been put forward to explain the formation history of GW231123.
These include the merger of the remnant black holes of two massive, low-metallicity stars with magnetic fields \citep[][]{gottlieb_spinning_2025}; formation from two massive highly spinning stars \citep[][]{croon_can_2025}; the merger of two primordial black holes \citep[][]{luca_gw231123_2025, yuan_gw231123_2025}; high spin due to gas accretion \citep[][]{bartos_accretion_2025, kiroglu_beyond_2025}; the merger of population III stars \citep[][]{liu_formation_2025, tanikawa_gw231123_2025}; and a hierarchical merger in an AGN disk \citep[][]{delfavero_prospects_2025} or in a cluster \citep[][]{paiella_assembling_2025}. 

Two further effects potentially complicate interpretation of GW231123: (i) waveform-model systematics at high spins and (ii) non-stationary detector noise. 
The importance of waveform systematics is highlighted by the discrepant results obtained with different waveforms in \cite{collaboration_gw231123_2025}.
The \textsc{NRSur7dq4} \citep{varma_surrogate_2019} waveform yields posterior distributions consistent with $\chi_1=\chi_2=1$, while \textsc{IMRPhenomXO4a} \citep{thompson_phenomxo4a_2024} yields posteriors consistent with $\chi_1=1$ but $\chi_2=0$. 
As gravitational waveform models are not in general calibrated for $\chi_{1,2}>0.8$, large systematics for GW231123 are not unexpected. 
\textsc{IMRPhenomXO4a} \citep{thompson_phenomxo4a_2024}~ is favored over the \textsc{NRSur7dq4}\footnote{Henceforth, we abbreviate \textsc{IMRPhenomXO4a} as \textsc{XO4a} and \textsc{NRSur7dq4} as \textsc{NRSur}.} \citep{varma_surrogate_2019} waveform with a Bayes factor of 200~\citep[see Appendix B of][]{collaboration_gw231123_2025}. 
While the Bayes factor must be balanced with our prior beliefs about model accuracy \citep[e.g.,][]{hoy_accelerating_2022, hoy_incorporation_2025} --- that is, the better fit for \textsc{XO4a} must be balanced against our prior belief in the better accuracy of \textsc{NRSur} --- a preference against a model better calibrated to more completely capture the physics of the merger is concerning.

In addition to the above, short signals like GW231123 contain only a few cycles, which make inferences particularly sensitive to non-stationary noise~\citep{miller, Udall:2024ovp}. 
A systematics study performed in \citet[][]{collaboration_gw231123_2025} found it difficult to replicate the degree of waveform-model disagreement observed in GW231123 with numerical-relativity injections, which may indicate that the underlying noise model is misspecified \citep{wmf}.
Glitches were detected in both the LIGO Hanford (H1) and the LIGO Livingston (L1) detector data close to the merger time of GW231123 \citep[][]{collaboration_gw231123_2025}; the extent to which data quality issues affect parameter inference remains unclear, but one should exercise caution. 

In this work, we focus on the hierarchical merger scenario as an explanation for GW231123. This scenario assumes that dense stellar environments harbor hierarchical mergers of compact objects \citep[e.g.,][]{rodriguez_black_2019, fragione_origin_2020, arca-sedda_breaching_2021, dallamico_gw190521_2021, kimball_evidence_2021, liu_hierarchical_2021, mapelli_hierarchical_2021, arcasedda_isolated_2023, torniamenti_hierarchical_2024, mahapatra_reconstructing_2024-1, antonini_star_2025, borchers_gravitational-wave_2025, mahapatra_predictions_2025-1}. Despite the challenges accounting for waveform systematics and non-Gaussian noise, we take the inferred spins from \cite{collaboration_gw231123_2025} at face value.

We test whether an event with the spin properties of both components of GW231123 can be assembled in a cluster, focusing in particular on the merger of two second-generation black holes.~\footnote{In our test, we do \textit{not} include information about whether the masses of the black holes in GW231123 can be plausibly created in a cluster as mass distributions can be much more uncertain}. 
To carry out this test, we employ the method described in \citet[][]{passengerAreAllModels2024}, which provides a framework for testing the ability of a population model to produce an exceptional event. 
We consider results from both the \textsc{NRSur} and \textsc{XO4a} waveform models. We find that it is possible to produce an event with component spins like GW231123 from our theoretical model, with $p\approx 5\%$.

\section{Method}
\subsection{Hierarchical-merger model}
We make use of the work of \citet[][]{borchers_gravitational-wave_2025}, who used the Cluster Monte Carlo Code~\citep[CMC; ][]{rodriguez_modeling_2022}, a software package which implements a star-by-star, orbit-averaged, Fokker-Planck Monte Carlo framework for modelling collisional dynamics in globular clusters. \citet[][]{borchers_gravitational-wave_2025} investigated the effect of black-hole post-merger kicks on the retained distribution of remnant spins $\chi_f$. 
This model assumes that the cumulative effect of two-body interactions in a cluster can be approximated as a sequence of pairwise interactions between nearest neighbors on a relaxation timescale, rather than by direct N-body integration of the full system.

\citet[][]{borchers_gravitational-wave_2025} found the $\chi_f$ distribution becomes more complex, broadening or becoming bimodal, depending on the spins of the progenitor black holes and the escape velocity of the cluster \citep[see also,][]{mahapatra_remnant_2021, araujo-alvarez_kicking_2024-1}. This bimodality arises primarily in clusters with low escape velocities, as systems that are either nearly aligned or anti-aligned receive weaker kicks on average and their remnants are therefore more commonly retained, than those with isotropically-distributed spins.

\begin{figure*}
    \centering
    \includegraphics[width=15cm]{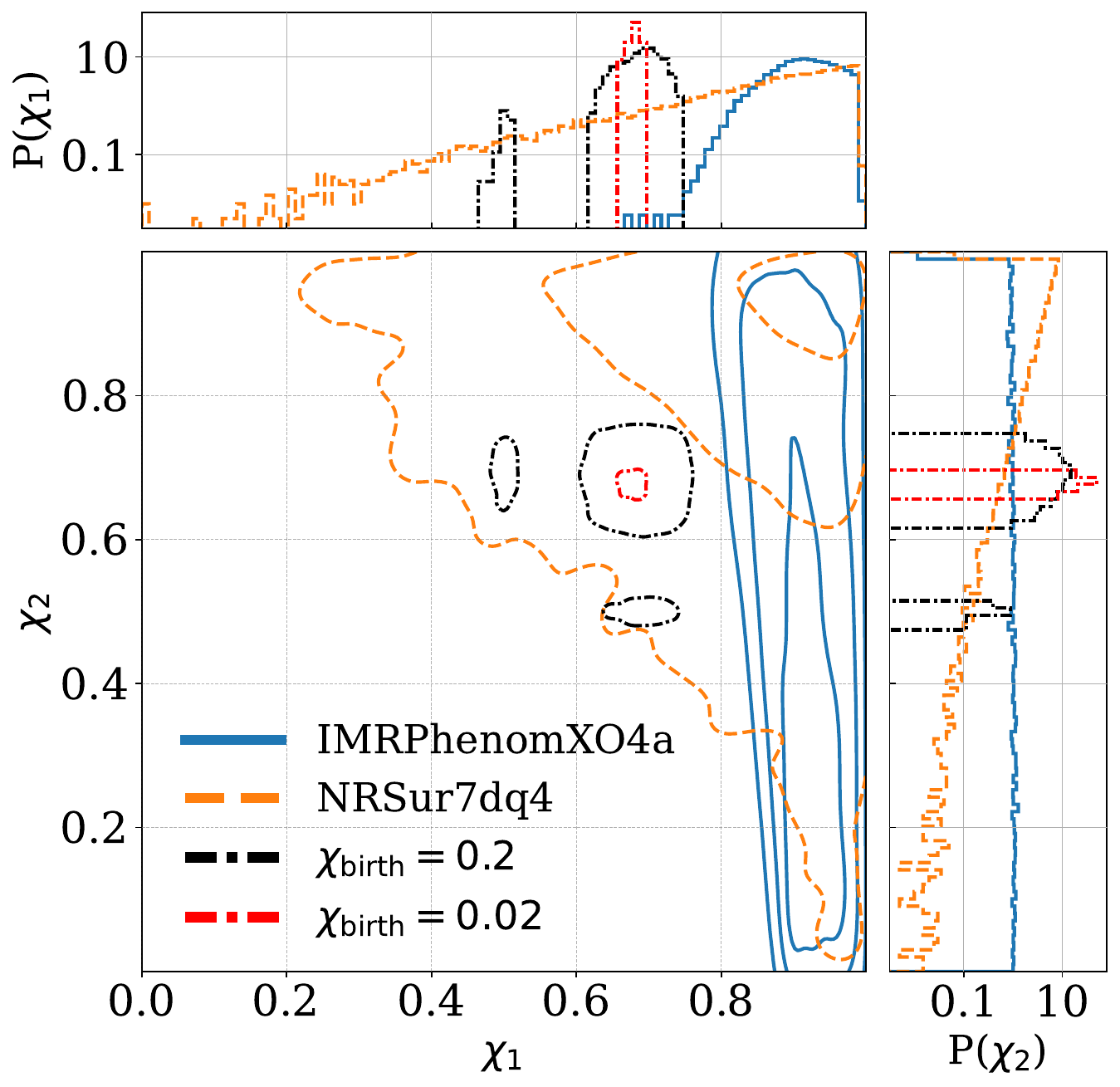}
    \caption{Comparing the predicted distribution of remnant black hole spins in globular clusters to the posteriors on $\chi_1, \chi_2$ for GW231123.
    In solid blue we plot the posterior for the \textsc{IMRPhenomXO4a} waveform model  and in dashed orange we plot the posterior for the \textsc{NRSur7dq4} waveform model.
    The three curves represent, one-, two-, and three-sigma intervals.
    The dotted curves show the distribution expected for globular clusters; $\chi_\text{birth} = 0.2$ in black and $\chi_\text{birth} = 0.02$ in red. 
    These contours are 99\% credible intervals. 
    Note that the probability axes in the 1D distributions are plotted with a log scale.}
    \label{fig:XO4a_and_NRSur_and_globular_cluster}
\end{figure*}

From \citet[][]{borchers_gravitational-wave_2025}, we use the distribution of $\chi_f$ for second-generation black holes retained by the cluster; see Fig.~\ref{fig:XO4a_and_NRSur_and_globular_cluster}.
The distribution is marginalised over the cluster escape velocity and mass-ratio distributions considered in \citet[][]{borchers_gravitational-wave_2025} and \citet[][]{rodriguez_modeling_2022}. We note that the initial mass distribution of globular clusters is uncertain, which directly impacts the escape velocity distribution, and thus our conclusions will mostly be relevant for typical escape velocities assumed in simulations. We assume that natal black holes have $\chi_\text{birth}=0.2$~\citep[consistent with most of the binary black hole detections; ][]{both_spin}. 
This choice is intended to be a conservative estimate of the birth spins of black holes, allowing for the possibility that angular momentum transport might not be efficient in high-mass stars.

\subsection{Formalism}
\citet[][]{passengerAreAllModels2024} derived a statistical formalism to test if an extremal value measured for some parameter is consistent with a population model. In order to explain the basic idea of \cite{passengerAreAllModels2024}, it is useful to consider the zero-noise limit where the binary parameters are measured perfectly.
We seek to determine if the measured value of some extreme parameter of interest $\widehat{x}$ is consistent with our population model for the distribution $\pi_\text{pop}(x)$.
If our measurement $\widehat{x}$ is noise-free, then we can calculate a $p$-value by just finding the probability mass such that \citep[e.g.,][]{fishbach_most_2020}
\begin{align}\label{eq:original_p_value}
    p = \int_{x \ge \widehat{x}_\text{event}} dx \, \pi_\text{pop}(x) .
\end{align}
A small $p$-value implies that the population model $\pi_\text{pop}(x)$ does not provide a good explanation of the measurement $\widehat{x}$.
Now, when the assumption about zero-noise is relaxed, the noise statistics are captured by the likelihood function. We employ the traditional Whittle likelihood, which assumes that noise is colored Gaussian~\citep[e.g.,][]{whittle_analysis_1953, cornish_towards_2013, thrane_introduction_2019},
\begin{align}
    \mathcal{L}(\tilde{d} | \theta)
    &= \prod_{j} \frac{2 \Delta f}{\pi P_j}\,
    \exp\!\left(-2\Delta f\,\frac{\lvert \tilde{d}_j - \tilde{\mu}_j(\theta)\rvert^2}{P_j}\right) , 
\end{align}
where $\tilde{d}$ is the frequency-domain strain data, $P_j$ is the single-sided noise power spectral density in the $j$-th frequency bin, $\Delta f$ is the frequency spacing and $\tilde{\mu}_j$ is the gravitational-waveform model (in the $j$-th bin) evaluated for binary parameters $\theta$.

To calculate a $p$-value, we first define a statistic that we call the `normalised evidence'~\citep[see][]{passengerAreAllModels2024}, 
\begin{align}\label{eq:original_normalised_evidence}
    \overline{\mathcal{Z}} \equiv & 
    \frac{ 
    \int d\theta \, {\mathcal{L}}(\tilde{d} | \theta) \pi_\text{pop}(x)\pi( \eta )
    }{ 
    \int d\theta \, {\mathcal{L}}(\tilde{d} | \theta)
    \pi_\text{U}(x)\pi( \eta)
    } .
\end{align}
Here, $x$ represents the parameter of interest with extremal values, $\eta$ represents the remaining binary parameters and the subscript $U$ represents a uniform prior. This quantity is effectively a Bayes factor comparing the evidence for a signal being drawn from the population model $\pi_\text{pop}(x)$ against the evidence for the signal being drawn from a fiducial uniform prior $\pi_\text{U}(x)$ on the parameter of interest. The denominator serves to `normalise' $\overline{\mathcal{Z}}$ by removing any dependence on parameters other than the parameter of interest~\citep[see][for more details]{passengerAreAllModels2024}.

To calculate our $p$-value (as in Eq. \ref{eq:original_p_value}), we find the probability mass of $\overline{\mathcal{Z}}$ for a simulated population, $\pi_\text{pop}(\overline{\mathcal{Z}})$, below the $\mathcal{Z}_\text{event}$ for a detected event,
\begin{align}\label{eq:original_Z_p_value}
    p = \int_{\overline{\mathcal{Z}} \le \overline{\mathcal{Z}}_\text{event}} d\overline{\mathcal{Z}} \, \pi_\text{pop}(\overline{\mathcal{Z}}) .
\end{align}
The specific question we want to answer is: are the large spins of GW231123 consistent with the hypothesis that it is produced from the merger of two second-generation black holes in a globular cluster?
Our astrophysical model is a distribution for the component spin magnitudes $(\chi_1, \chi_2)$ of 2G+2G black holes, $\pi_\text{2G}(\chi_1, \chi_2)$ obtained by \cite{borchers_gravitational-wave_2025} for birth spins of $\chi_{\rm birth} = 0.2$.
We choose priors on the remaining binary parameters $\eta$ to be the same as those chosen for the LVK's analysis of GW231123 \citep[][]{collaboration_gw231123_2025} such that our combined set of priors can be written as $\pi_\text{2G}(\chi_1, \chi_2)\, \pi( \eta)$.

Following \citet[][]{passengerAreAllModels2024}, we wish to calculate a $p$-value to quantify the probability that GW231123 has spin values drawn from $\pi_\text{2G}(\chi_1, \chi_2)$. 
As in Eq. \ref{eq:original_normalised_evidence}, we define the normalised evidence as
\begin{align}\label{eq:normalised_evidence}
    \overline{\mathcal{Z}} \equiv & 
    \frac{ 
    \int d\theta \, {\mathcal{L}}(\tilde{d} | \theta) \pi_\text{2G}(\chi_1, \chi_2)\pi( \eta )
    }{ 
    \int d\theta \, {\mathcal{L}}(\tilde{d} | \theta)
    \pi_\text{U}(\chi_1, \chi_2)\pi( \eta)
    } .
\end{align}
Next, we simulate events drawn from $\pi_\text{2G}(\chi_1, \chi_2)\pi( \eta )$ and inject them into Gaussian noise with power spectral density $P$, while accounting for selection effects.
For each event, we calculate $\overline{\mathcal{Z}}$ to produce an empirical distribution $\pi_\text{pop}(\overline{\mathcal{Z}})$.
Using Eq. \ref{eq:original_Z_p_value}, we calculate a $p$-value by comparing the normalized evidence for GW231123 with the distribution of normalised evidence values for signals drawn from the astrophysical population.

\subsection{Application to \ac{GW} data}
We perform two analyses: one using the \textsc{XO4a} waveform, as it is the waveform most favoured in explaining GW231123 by way of Bayes factor, and \textsc{NRSur}, as it is the waveform most closely calibrated to numerical relativity.
To simulate a hierarchical-merger distribution of \ac{GW} events, we adopt the following procedure:
\begin{enumerate}
    \item For each injection, we assign component spins $(\chi_1, \chi_2)$ from the $\chi_f$ distribution with $\chi_\text{birth}=0.2$ from \citet[][]{borchers_gravitational-wave_2025}, marginalised over cluster escape velocity.
    We use this birth spin to make it \textit{easier} for the cluster model to make events with large component spins.
    This way, if we obtain a small $p$-value, we know it would only be smaller had we used smaller birth spins.
    \item We assign component masses from the posterior distributions of either the \textsc{XO4a} or \textsc{NRSur} analysis of GW231123. We assign the remaining extrinsic parameters for each injection by randomly sampling from standard LVK priors \citep[e.g.,][]{collaboration_gwtc-1_2019, abbott_gwtc-2_2021, ligoscientificcollaborationGWTC3CompactBinary2023}. We account for selection effects and ensure each event has a network optimal signal-to-noise ratio $>{12}$.
    \item For both \textsc{XO4a} and \textsc{NRSur}, we use the following priors during analysis: we sample chirp mass $\mathcal{M}$ uniformly in the range $[30,200]\text{M}_\odot$. For \textsc{XO4a}, we sample the mass ratio $q$ uniformly in the range $[1/10,1]$, while for \textsc{NRSur}, we sample it uniformly in the range $[1/6,1]$. This reduced mass range is due to limitations of the \textsc{NRSur} model \citep[][]{varma_surrogate_2019}. The priors on the remaining extrinsic parameters for each injection follow standard LVK analysis priors. 
    \item We inject each simulated signal into a $\unit[2]{s}$ segment of Gaussian noise colored by the power spectral density used in the public analysis of GW231123. We analyse these signals using both the \textsc{XO4a} and \textsc{NRSur} waveform models, using the \textsc{dynesty} nested sampler \citep[][]{speagleDynestyDynamicNested2020a} implemented in the \textsc{Bilby} software package \citep[][]{ashtonBilbyUserfriendlyBayesian2019, romero-shaw_bayesian_2020}. For \textsc{XO4a} we analyse $119$ simulated signals and for \textsc{NRSur} we analyse $97$ simulated signals. With this many injections, we can determine if GW231123 is unusual with a scarcity of around $1\%$, but we are unable to distinguish between $p$-values of $\sim{1\%}$ and five-sigma outliers with $p=3.5\times10^{-6}$. We perform two calculations: one in which the prior on $(\chi_1, \chi_2)$ is $\pi_\text{2G}(\chi_1, \chi_2)$ and one in which it is uniform in the range [0,1]; i.e., $\pi_\text{U}(\chi_1, \chi_2)$.
    \item We analyse GW231123 using $\unit[2]{s}$ of data centered around the trigger time of the event, using the same settings as above. Following \citet[][]{collaboration_gw231123_2025}, we mitigate against the glitch in the H1 detector using the \textsc{BayesWave} \citep[][]{cornishBayeswaveBayesianInference2015} software package. We perform separate analyses using the \textsc{XO4a} and \textsc{NRSur} waveform models.
    \item We calculate $\overline{\mathcal{Z}}$ for each signal (simulated and GW231123) and each waveform model using Eq.~\ref{eq:normalised_evidence}.
\end{enumerate}

\section{Results \& Conclusion}\label{results}
\begin{figure*}
    \centering
    \includegraphics[width=180mm]{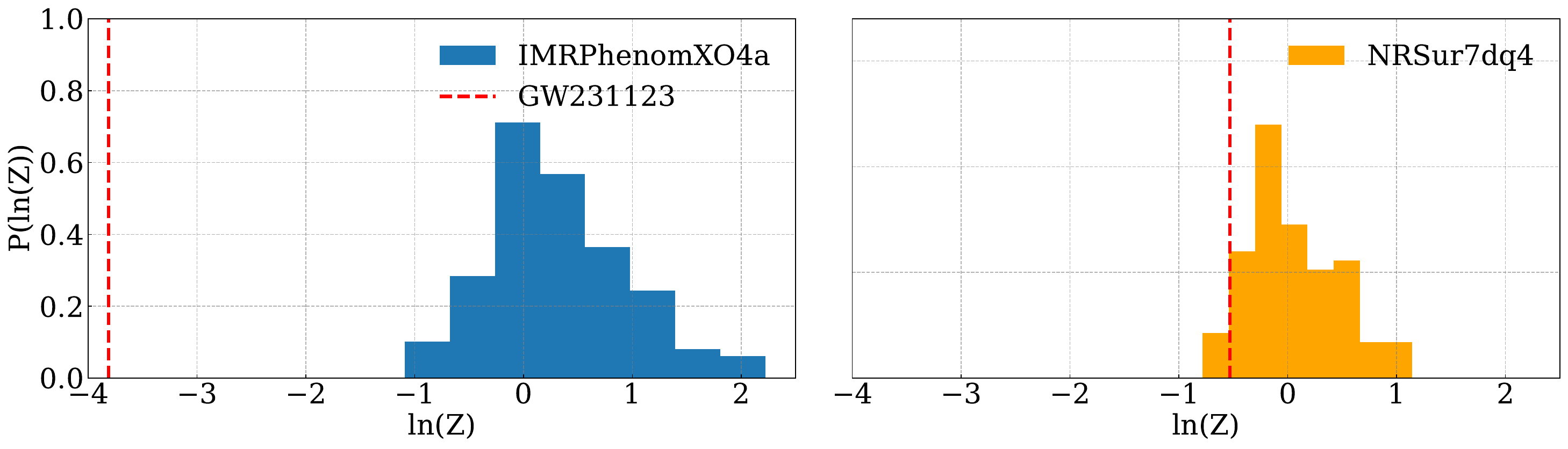}
    \caption{Distributions of normalized evidence $\overline{\mathcal{Z}}$ from \cite{passengerAreAllModels2024}.
    The histograms show the distribution predicted by our cluster model, while the dashed red lines show the value measured for GW231123.
    For the \textsc{XO4a} analysis, GW231123 is excluded from the injected distribution with $p\ll{1\%}$. For the \textsc{NRSur} analysis, GW231123 is excluded with $p \approx 5\%$.}
    \label{fig:ln_Z_histogram}
\end{figure*}

In Fig.~\ref{fig:ln_Z_histogram}, we show the empirical distribution $\pi_\text{pop}(\overline{\mathcal{Z}})$ for simulated hierarchical mergers compared to the $\overline{\mathcal{Z}}$ calculated for GW231123.
The results for \textsc{XO4a} are shown on the left while the results for \textsc{NRSur} are shown on the right. 
Using \textsc{XO4a}, GW231123 is excluded from the distribution of $\pi_\text{pop}(\overline{\mathcal{Z}})$ for simulated hierarchical mergers, with $p\ll{1\%}$ (Equation \ref{eq:original_Z_p_value}). 
In contrast, for the \textsc{NRSur} waveform model, we obtain $p \approx 5\%$. Thus, given our assumptions and the remnant-spin model from \citet{borchers_gravitational-wave_2025}, we reject the 2G+2G hierarchical formation channel for GW231123 when the data are analyzed with \textsc{XO4a}, but we find there is a reasonable chance that GW231123 is a 2G+2G hierarchical merger when we analyze the data with \textsc{NRSur}. While we think it is unlikely these conclusions would change drastically under different models of 2G+2G mergers in clusters, it is possible the $p$-values could change somewhat.

For a better qualitative understanding of this result, we return to Fig.~\ref{fig:XO4a_and_NRSur_and_globular_cluster}.
We see that, while \textsc{NRSur} prefers that both black holes have large spins, it allows for $\chi_1$ to be smaller than \textsc{XO4a}, which requires $\chi_1\gtrsim0.8$.
This makes it possible (though somewhat unlikely) that GW231123 has spins of $\chi_1 \approx \chi_2 \approx 0.8$, which are mildly consistent both with our population model and also with the GW231123 posterior.

The compatibility of GW231123 with our model for 2G+2G mergers is waveform dependent, as different waveform models give conflicting spin-magnitude estimates. This demonstrates the necessity for the development of waveform models that are well calibrated against numerical relativity in this part of the parameter space.

While we mitigate against the effect of the glitch detected in H1, it remains possible that inference of the spins of GW231123 is contaminated by non-Gaussian noise. A recent study from \citet[][]{ray_gw231123_2025} also showed that microglitches may bias spin measurements in events like GW231123 to be more consistent with $\chi\approx1$. Development of more flexible noise models, especially those that explicitly model glitches, may improve the reliability of GW231123’s spin inference.

We do not consider how the extremal component masses of GW231123 ($m_1=137^{+22}_{-17}$ $\text{M}_\odot$ and $m_2=103^{+20}_{-52}$ $\text{M}_\odot$) affect interpretation of it as a 2G+2G hierarchical merger. Doing so would require simultaneous modelling of the mass distribution of black holes in 2G+2G mergers, which are not considered in the model for globular clusters developed by~\citet[][]{borchers_gravitational-wave_2025}. 
At any rate, we expect the result may depend strongly on the assumed distribution of 1G black hole mass \citep[e.g.,][]{kimball_what_2020-1} and it is therefore unlikely to be much more informative as their distribution will be much more uncertain.

Our results suggest that it is possible for an event like GW231123 to have originated through a hierarchical merger scenario. This opens the door for other alternate interpretations~\citep[e.g.,][]{gottlieb_spinning_2025, croon_can_2025, luca_gw231123_2025, yuan_gw231123_2025, bartos_accretion_2025, kiroglu_beyond_2025, tanikawa_gw231123_2025, delfavero_prospects_2025}. Most of these formation scenarios come with a higher degree of uncertainty than hierarchical mergers, sometimes with poorly understood physics. If either GW231123 or other events like this in the future are found to be more closely associated with one of these alternative formation scenarios, they can provide insight into the physics governing such scenarios. 
 
While we focus here on the possibility that GW231123 is a 2G+2G merger, one may also consider the possibility that this event is a 2G+1G or 3G+1G merger.
In fact, parameter estimation of GW231123 using the \textsc{XO4a} waveform gives component mass estimates $m_1=143^{+24}_{-14}$ $\text{M}_\odot$ and $m_2=55^{+11}_{-18}$ $\text{M}_\odot$, and component spin estimates $\chi_1=0.92^{+0.07}_{-0.06}$ and $\chi_2=0.47^{+0.41}_{-0.47}$ \citep[][] {collaboration_gw231123_2025}, which may be more consistent with the merger of a high-mass, high-spin second or third-generation black hole with a first-generation black hole. Again, a comprehensive treatment of GW231123 as a 2G+1G or 3G+1G merger would likely be more sensitive to the assumed mass distribution of 1G black holes -- we leave this for future work.

Moreover, GW231123 may also be explained as a 3G+2G hierarchical merger, due especially to the extremal spin of its primary component estimated by all waveforms. For example, \citet[][]{borchers_gravitational-wave_2025} find that the retained remnant spin distribution of 2G+1G mergers (3G components) form consistently with $\chi_f\ge{0.8}$, even at low cluster escape velocities and birth spins. However, we suspect that the merger rates of 3G+2G events are probably too low \citep[compared to 2G+2G or 2G+1G mergers; see][]{kimball_black_2020,gerosa_hierarchical_2021} to account for GW231123. Such mergers may require alternate formation environments with higher escape velocities, such as nuclear star clusters \citep[e.g.,][]{fragione_demographics_2023}. The possibility of GW231123 containing a 3G or higher generation was also raised as a possibility in \citet[][]{collaboration_gw231123_2025}.

\section*{Data Availability}

The data underlying this article are publicly available at \url{https://www.gw-openscience.org}.

\section*{Acknowledgements}
This is LIGO document \#P2500568.
We acknowledge support from the Australian Research
Council (ARC) Centres of Excellence CE170100004 and CE230100016, as well as ARC
LE210100002, and ARC DP230103088. This research was supported by the Commonwealth through an Australian Government Research Training Program Scholarship [DOI: https://doi.org/10.82133/C42F-K220].
This material is based upon work supported by NSF’s LIGO Laboratory
which is a major facility fully funded by the National
Science Foundation. The authors are grateful for computational resources provided by the LIGO Laboratory
and supported by National Science Foundation Grants
PHY-0757058 and PHY-0823459.

This research has made use of data or software obtained from the Gravitational Wave Open Science Center (gw-openscience.org), a service of LIGO Laboratory,
the LIGO Scientific Collaboration, the Virgo Collaboration, and KAGRA. LIGO Laboratory and Advanced
LIGO are funded by the United States National Science Foundation (NSF) as well as the Science and Technology Facilities Council (STFC) of the United Kingdom, the Max-Planck-Society (MPS), and the State of
Niedersachsen/Germany for support of the construction
of Advanced LIGO and construction and operation of
the GEO600 detector. Additional support for Advanced
LIGO was provided by the Australian Research Council.
Virgo is funded, through the European Gravitational
Observatory (EGO), by the French Centre National
de Recherche Scientifique (CNRS), the Italian Istituto
Nazionale di Fisica Nucleare (INFN) and the Dutch
Nikhef, with contributions by institutions from Belgium,
Germany, Greece, Hungary, Ireland, Japan, Monaco,
Poland, Portugal, Spain. The construction and operation of KAGRA are funded by Ministry of Education,
Culture, Sports, Science and Technology (MEXT), and
Japan Society for the Promotion of Science (JSPS), National Research Foundation (NRF) and Ministry of Science and ICT (MSIT) in Korea, Academia Sinica (AS)
and the Ministry of Science and Technology (MoST) in
Taiwan.
\bibliographystyle{aasjournal}
\bibliography{refs}

@String{apj = "Astrophys. J."}

@String{apjl = "Astrophys. J. Lett."}

@String{apjs = "Astrophys. J. Supp."}

@String{pasa = "Pub. Astron. Soc. Aust."}

@String{prd = "Phys. Rev. D"}

@article{wmf,
author = "I. M. Romero-Shaw and Eric Thrane and Paul D. Lasky",
title = "When models fail: an introduction to posterior predictive checks and model misspecification in gravitational-wave astronomy",
journal = pasa,
volume = 39,
pages = "E025",
year = 2022}

@article{miller,
author = "Simona J. Miller
and Maximiliano Isi
and Katerina Chatziioannou
and Vijay Varma and Ilya Mandel",
journal = prd,
volume = 109,
pages = "024024",
year = 2024}

@article{gwtc-4_pop,
author = "Abac and others",
title = "{GWTC-4.0: Population Properties of Merging Compact Binaries}",
journal = "",
volume = "",
pages = "",
year = 2025,
note = "arxiv2508.18083"}

@article{gwtc-3_hui,
author = "H. Tong and S. Galaudage and Eric Thrane",
title = "{The population properties of spinning black holes using Gravitational-wave Transient Catalog 3}",
journal = prd,
volume = 106,
pages = "103019",
year = 2022}

@article{both_spin,
author = "C. Adamcewicz and N. Guttman and Paul D. Lasky and Eric Thrane",
title = "Do both black holes spin in merging binaries? Evidence from GWTC-4 and astrophysical implications",
journal = "",
volume = "",
pages = "",
year = 2025,
note = "arxiv/2509.04706"}

@article{single-spin,
author = "C. Adamcewicz and Paul D. Lasky and Eric Thrane",
title = "Which black hole is spinning? Probing the origin of black-hole spin with gravitational waves",
journal = apjl,
volume = 964,
pages = "L6",
year = 2024}

@article{passengerAreAllModels2024,
	title = {Are all models wrong? {Falsifying} binary formation models in gravitational-wave astronomy using exceptional events},
	volume = {535},
	copyright = {https://creativecommons.org/licenses/by/4.0/},
	issn = {0035-8711, 1365-2966},
	shorttitle = {Are all models wrong?},
	url = {https://academic.oup.com/mnras/article/535/3/2837/7890803},
	doi = {10.1093/mnras/stae2521},
	language = {en},
	number = {3},
	urldate = {2025-06-18},
	journal = {Monthly Notices of the Royal Astronomical Society},
	author = {Passenger, Lachlan and Thrane, Eric and Lasky, Paul and Payne, Ethan and Stevenson, Simon and Farr, Ben},
	month = nov,
	year = {2024},
	pages = {2837--2843},
}

@article{theligoscientificcollaborationAdvancedLIGO2015,
	title = {Advanced {LIGO}},
	volume = {32},
	issn = {0264-9381, 1361-6382},
	url = {http://arxiv.org/abs/1411.4547},
	doi = {10.1088/0264-9381/32/7/074001},
	number = {7},
	urldate = {2024-01-16},
	journal = {Class. Quantum Grav.},
	author = {Aasi et al., J},
	month = apr,
	year = {2015},
	note = {arXiv:1411.4547 [astro-ph, physics:gr-qc, physics:physics]},
	pages = {074001},
}

@article{acerneseAdvancedVirgo2nd2015a,
	title = {Advanced {Virgo}: a 2nd generation interferometric gravitational wave detector},
	volume = {32},
	issn = {0264-9381, 1361-6382},
	shorttitle = {Advanced {Virgo}},
	url = {http://arxiv.org/abs/1408.3978},
	doi = {10.1088/0264-9381/32/2/024001},
	number = {2},
	urldate = {2024-01-16},
	journal = {Class. Quantum Grav.},
	author = {Acernese, F. and Agathos, M. and Agatsuma, K. and Aisa, D. and Astone, P. and Ballardin, G. and Barone, F. and Baronick, J.-P. and Barsuglia, M. and Basti, A. and Basti, F. and Bauer, Th S. and Bavigadda, V. and Bejger, M. and Beker, M. G. and Belczynski, C. and Bertolini, A. and Bitossi, M. and Bizouard, M. A. and Bloemen, S. and Blom, M. and Boer, M. and Bogaert, G. and Bondi, D. and Bondu, F. and Bonelli, L. and Bonnand, R. and Boschi, V. and Bosi, L. and Bouedo, T. and Bradaschia, C. and Branchesi, M. and Briant, T. and Bulik, T. and Bulten, H. J. and Buskulic, D. and Buy, C. and Cagnoli, G. and Calloni, E. and Canuel, B. and Carbognani, F. and Cavalier, F. and Cavalieri, R. and Cella, G. and Cesarini, E. and Chassande-Mottin, E. and Chincarini, A. and Chiummo, A. and Chua, S. and Cleva, F. and Coccia, E. and Cohadon, P.-F. and Colla, A. and Colombini, M. and Conte, A. and Coulon, J.-P. and Cuoco, E. and Dalmaz, A. and D'Antonio, S. and Dattilo, V. and Davier, M. and Day, R. and Debreczeni, G. and Degallaix, J. and Deléglise, S. and Del Pozzo, W. and Dereli, H. and De Rosa, R. and Di Fiore, L. and Di Lieto, A. and Doets, M. and Dolique, V. and Drago, M. and Ducrot, M. and Endrőczi, G. and Fafone, V. and Farinon, S. and Ferrante, I. and Ferrini, F. and Fidecaro, F. and Fiori, I. and Flaminio, R. and Fournier, J.-D. and Franco, S. and Frasca, S. and Frasconi, F. and Gammaitoni, L. and Garufi, F. and Gaspard, M. and Gemme, G. and Gendre, B. and Genin, E. and Gennai, A. and Ghosh, S. and Giacobone, L. and Giazotto, A. and Gouaty, R. and Granata, M. and Greco, G. and Groot, P. and Guidi, G. M. and Harms, J. and Heidmann, A. and Heitmann, H. and Hello, P. and Hemming, G. and Hennes, E. and Hofman, D. and Jaranowski, P. and Jonker, R. J. G. and Kasprzack, M. and Kéfélian, F. and Kowalska, I. and Kraan, M. and Królak, A. and Kutynia, A. and Lazzaro, C. and Leonardi, M. and Leroy, N. and Letendre, N. and Li, T. G. F. and Lieunard, B. and Loriette, V. and Losurdo, G. and Magazzù, C. and Majorana, E. and Maksimovic, I. and Malvezzi, V. and Man, N. and Mangano, V. and Mantovani, M. and Marchesoni, F. and Marion, F. and Marque, J. and Martelli, F. and Martellini, L. and Masserot, A. and Meacher, D. and Meidam, J. and Mezzani, F. and Michel, C. and Milano, L. and Minenkov, Y. and Moggi, A. and Mohan, M. and Montani, M. and Morgado, N. and Mours, B. and Mul, F. and Nagy, M. F. and Nardecchia, I. and Naticchioni, L. and Nelemans, G. and Neri, I. and Neri, M. and Nocera, F. and Pacaud, E. and Palomba, C. and Paoletti, F. and Paoli, A. and Pasqualetti, A. and Passaquieti, R. and Passuello, D. and Perciballi, M. and Petit, S. and Pichot, M. and Piergiovanni, F. and Pillant, G. and Piluso, A. and Pinard, L. and Poggiani, R. and Prijatelj, M. and Prodi, G. A. and Punturo, M. and Puppo, P. and Rabeling, D. S. and Rácz, I. and Rapagnani, P. and Razzano, M. and Re, V. and Regimbau, T. and Ricci, F. and Robinet, F. and Rocchi, A. and Rolland, L. and Romano, R. and Rosińska, D. and Ruggi, P. and Saracco, E. and Sassolas, B. and Schimmel, F. and Sentenac, D. and Sequino, V. and Shah, S. and Siellez, K. and Straniero, N. and Swinkels, B. and Tacca, M. and Tonelli, M. and Travasso, F. and Turconi, M. and Vajente, G. and van Bakel, N. and van Beuzekom, M. and Brand, J. F. J. van den and Broeck, C. Van Den and van der Sluys, M. V. and van Heijningen, J. and Vasúth, M. and Vedovato, G. and Veitch, J. and Verkindt, D. and Vetrano, F. and Viceré, A. and Vinet, J.-Y. and Visser, G. and Vocca, H. and Ward, R. and Was, M. and Wei, L.-W. and Yvert, M. and Zadrożny, A. and Zendri, J.-P.},
	month = jan,
	year = {2015},
	note = {arXiv:1408.3978 [gr-qc, physics:physics]},
	pages = {024001},
}

@article{akutsuKAGRAGenerationInterferometric2019,
	title = {{KAGRA}: 2.5 {Generation} {Interferometric} {Gravitational} {Wave} {Detector}},
	volume = {3},
	issn = {2397-3366},
	shorttitle = {{KAGRA}},
	url = {http://arxiv.org/abs/1811.08079},
	doi = {10.1038/s41550-018-0658-y},
	number = {1},
	urldate = {2024-01-16},
	journal = {Nat Astron},
	author = {Akutsu, T. and Ando, M. and Arai, K. and Arai, Y. and Araki, S. and Araya, A. and Aritomi, N. and Asada, H. and Aso, Y. and Atsuta, S. and Awai, K. and Bae, S. and Baiotti, L. and Barton, M. A. and Cannon, K. and Capocasa, E. and Chen, C.-S. and Chiu, T.-W. and Cho, K. and Chu, Y.-K. and Craig, K. and Creus, W. and Doi, K. and Eda, K. and Enomoto, Y. and Flaminio, R. and Fujii, Y. and Fujimoto, M.-K. and Fukunaga, M. and Fukushima, M. and Furuhata, T. and Haino, S. and Hasegawa, K. and Hashino, K. and Hayama, K. and Hirobayashi, S. and Hirose, E. and Hsieh, B. H. and Huang, C.-Z. and Ikenoue, B. and Inoue, Y. and Ioka, K. and Itoh, Y. and Izumi, K. and Kaji, T. and Kajita, T. and Kakizaki, M. and Kamiizumi, M. and Kanbara, S. and Kanda, N. and Kanemura, S. and Kaneyama, M. and Kang, G. and Kasuya, J. and Kataoka, Y. and Kawai, N. and Kawamura, S. and Kawasaki, T. and Kim, C. and Kim, J. and Kim, J. C. and Kim, W. S. and Kim, Y.-M. and Kimura, N. and Kinugawa, T. and Kirii, S. and Kitaoka, Y. and Kitazawa, H. and Kojima, Y. and Kokeyama, K. and Komori, K. and Kong, A. K. H. and Kotake, K. and Kozu, R. and Kumar, R. and Kuo, H.-S. and Kuroyanagi, S. and Lee, H. K. and Lee, H. M. and Lee, H. W. and Leonardi, M. and Lin, C.-Y. and Lin, F.-L. and Liu, G. C. and Liu, Y. and Majorana, E. and Mano, S. and Marchio, M. and Matsui, T. and Matsushima, F. and Michimura, Y. and Mio, N. and Miyakawa, O. and Miyamoto, A. and Miyamoto, T. and Miyo, K. and Miyoki, S. and Morii, W. and Morisaki, S. and Moriwaki, Y. and Morozumi, T. and Musha, M. and Nagano, K. and Nagano, S. and Nakamura, K. and Nakamura, T. and Nakano, H. and Nakano, M. and Nakao, K. and Narikawa, T. and Naticchioni, L. and Quynh, L. Nguyen and Ni, W.-T. and Nishizawa, A. and Ochi, T. and Oh, J. J. and Oh, S. H. and Ohashi, M. and Ohishi, N. and Ohkawa, M. and Okutomi, K. and Ono, K. and Oohara, K. and Ooi, C. P. and Pan, S.-S. and Park, J. and Arellano, F. E. Peña and Pinto, I. and Sago, N. and Saijo, M. and Saito, Y. and Sakai, K. and Sakai, Y. and Sakai, Y. and Sasai, M. and Sasaki, M. and Sasaki, Y. and Sato, S. and Sato, T. and Sekiguchi, Y. and Seto, N. and Shibata, M. and Shimoda, T. and Shinkai, H. and Shishido, T. and Shoda, A. and Somiya, K. and Son, E. J. and Suemasa, A. and Suzuki, T. and Suzuki, T. and Tagoshi, H. and Tahara, H. and Takahashi, H. and Takahashi, R. and Takamori, A. and Takeda, H. and Tanaka, H. and Tanaka, K. and Tanaka, T. and Tanioka, S. and Martin, E. N. Tapia San and Tatsumi, D. and Tomaru, T. and Tomura, T. and Travasso, F. and Tsubono, K. and Tsuchida, S. and Uchikata, N. and Uchiyama, T. and Uehara, T. and Ueki, S. and Ueno, K. and Ushiba, T. and van Putten, M. H. P. M. and Vocca, H. and Wada, S. and Wakamatsu, T. and Watanabe, Y. and Xu, W.-R. and Yamada, T. and Yamamoto, A. and Yamamoto, K. and Yamamoto, K. and Yamamoto, S. and Yamamoto, T. and Yokogawa, K. and Yokoyama, J. and Yokozawa, T. and Yoon, T. H. and Yoshioka, T. and Yuzurihara, H. and Zeidler, S. and Zhu, Z.-H.},
	month = jan,
	year = {2019},
	note = {arXiv:1811.08079 [astro-ph, physics:gr-qc, physics:physics]},
	pages = {35--40},
}

@article{hegerNucleosyntheticSignaturePopulation2002,
	title = {The {Nucleosynthetic} {Signature} of {Population} {III}},
	volume = {567},
	issn = {0004-637X},
	url = {https://iopscience.iop.org/article/10.1086/338487/meta},
	doi = {10.1086/338487},
	language = {en},
	number = {1},
	urldate = {2024-01-24},
	journal = {ApJ},
	author = {Heger, A. and Woosley, S. E.},
	month = mar,
	year = {2002},
	note = {Publisher: IOP Publishing},
	pages = {532},
}

@article{woosleyPulsationalPairInstability2007,
	title = {Pulsational pair instability as an explanation for the most luminous supernovae},
	volume = {450},
	copyright = {2007 Springer Nature Limited},
	issn = {1476-4687},
	url = {https://www.nature.com/articles/nature06333},
	doi = {10.1038/nature06333},
	language = {en},
	number = {7168},
	urldate = {2024-01-24},
	journal = {Nature},
	author = {Woosley, S. E. and Blinnikov, S. and Heger, Alexander},
	month = nov,
	year = {2007},
	note = {Number: 7168
Publisher: Nature Publishing Group},
	pages = {390--392},
}

@article{belczynskiEffectPairinstabilityMass2016b,
	title = {The effect of pair-instability mass loss on black-hole mergers},
	volume = {594},
	copyright = {© ESO, 2016},
	issn = {0004-6361, 1432-0746},
	url = {https://www.aanda.org/articles/aa/abs/2016/10/aa28980-16/aa28980-16.html},
	doi = {10.1051/0004-6361/201628980},
	language = {en},
	urldate = {2024-01-24},
	journal = {A\&A},
	author = {Belczynski, K. and Heger, A. and Gladysz, W. and Ruiter, A. J. and Woosley, S. and Wiktorowicz, G. and Chen, H.-Y. and Bulik, T. and O’Shaughnessy, R. and Holz, D. E. and Fryer, C. L. and Berti, E.},
	month = oct,
	year = {2016},
	note = {Publisher: EDP Sciences},
	pages = {A97},
}

@article{woosleyPulsationalPairinstabilitySupernovae2017,
	title = {Pulsational {Pair}-instability {Supernovae}},
	volume = {836},
	issn = {0004-637X},
	url = {https://dx.doi.org/10.3847/1538-4357/836/2/244},
	doi = {10.3847/1538-4357/836/2/244},
	language = {en},
	number = {2},
	urldate = {2024-01-24},
	journal = {ApJ},
	author = {Woosley, S. E.},
	month = feb,
	year = {2017},
	note = {Publisher: The American Astronomical Society},
	pages = {244},
}

@article{woosleyEvolutionMassiveHelium2019,
	title = {The {Evolution} of {Massive} {Helium} {Stars}, {Including} {Mass} {Loss}},
	volume = {878},
	issn = {0004-637X},
	url = {https://dx.doi.org/10.3847/1538-4357/ab1b41},
	doi = {10.3847/1538-4357/ab1b41},
	language = {en},
	number = {1},
	urldate = {2024-01-24},
	journal = {ApJ},
	author = {Woosley, S. E.},
	month = jun,
	year = {2019},
	note = {Publisher: The American Astronomical Society},
	pages = {49},
}

@article{woosleyPairinstabilityMassGap2021,
	title = {The {Pair}-instability {Mass} {Gap} for {Black} {Holes}},
	volume = {912},
	issn = {2041-8205},
	url = {https://dx.doi.org/10.3847/2041-8213/abf2c4},
	doi = {10.3847/2041-8213/abf2c4},
	language = {en},
	number = {2},
	urldate = {2024-01-24},
	journal = {ApJL},
	author = {Woosley, S. E. and Heger, Alexander},
	month = may,
	year = {2021},
	note = {Publisher: The American Astronomical Society},
	pages = {L31},
}

@article{farmerMindGapLocation2019,
	title = {Mind the gap: {The} location of the lower edge of the pair instability supernovae black hole mass gap},
	volume = {887},
	issn = {0004-637X, 1538-4357},
	shorttitle = {Mind the gap},
	url = {http://arxiv.org/abs/1910.12874},
	doi = {10.3847/1538-4357/ab518b},
	number = {1},
	urldate = {2024-01-22},
	journal = {ApJ},
	author = {Farmer, R. and Renzo, M. and de Mink, S. E. and Marchant, P. and Justham, S.},
	month = dec,
	year = {2019},
	note = {arXiv:1910.12874 [astro-ph]},
	pages = {53},
}

@article{costaFormationGW190521Stellar2021,
	title = {Formation of {GW190521} from stellar evolution: the impact of the hydrogen-rich envelope, dredge-up, and {12C}(α, γ){16O} rate on the pair-instability black hole mass gap},
	volume = {501},
	issn = {0035-8711},
	shorttitle = {Formation of {GW190521} from stellar evolution},
	url = {https://doi.org/10.1093/mnras/staa3916},
	doi = {10.1093/mnras/staa3916},
	number = {3},
	urldate = {2024-01-22},
	journal = {Monthly Notices of the Royal Astronomical Society},
	author = {Costa, Guglielmo and Bressan, Alessandro and Mapelli, Michela and Marigo, Paola and Iorio, Giuliano and Spera, Mario},
	month = mar,
	year = {2021},
	pages = {4514--4533},
}

@article{winchPredictingHeaviestBlack2024,
	title = {Predicting the heaviest black holes below the pair instability gap},
	volume = {529},
	issn = {0035-8711},
	url = {https://ui.adsabs.harvard.edu/abs/2024MNRAS.529.2980W},
	doi = {10.1093/mnras/stae393},
	urldate = {2024-05-14},
	journal = {Monthly Notices of the Royal Astronomical Society},
	author = {Winch, Ethan R. J. and Vink, Jorick S. and Higgins, Erin R. and Sabhahitf, Gautham N.},
	month = apr,
	year = {2024},
	note = {Publisher: OUP
ADS Bibcode: 2024MNRAS.529.2980W},
	pages = {2980--3002},
}

@article{fuller_most_2019,
	title = {Most {Black} {Holes} are {Born} {Very} {Slowly} {Rotating}},
	volume = {881},
	issn = {2041-8205, 2041-8213},
	url = {http://arxiv.org/abs/1907.03714},
	doi = {10.3847/2041-8213/ab339b},
	number = {1},
	urldate = {2025-08-13},
	journal = {The Astrophysical Journal Letters},
	author = {Fuller, Jim and Ma, Linhao},
	month = aug,
	year = {2019},
	note = {arXiv:1907.03714 [astro-ph]},
	pages = {L1},
}

@article{rodriguez_modeling_2022,
	title = {Modeling {Dense} {Star} {Clusters} in the {Milky} {Way} and {Beyond} with the {Cluster} {Monte} {Carlo} {Code}},
	volume = {258},
	issn = {0067-0049, 1538-4365},
	url = {http://arxiv.org/abs/2106.02643},
	doi = {10.3847/1538-4365/ac2edf},
	number = {2},
	urldate = {2025-08-13},
	journal = {The Astrophysical Journal Supplement Series},
	author = {Rodriguez, Carl L. and Weatherford, Newlin C. and Coughlin, Scott C. and Seoane, Pau Amaro and Breivik, Katelyn and Chatterjee, Sourav and Fragione, Giacomo and Kıroğlu, Fulya and Kremer, Kyle and Rui, Nicholas Z. and Ye, Claire S. and Zevin, Michael and Rasio, Frederic A.},
	month = feb,
	year = {2022},
	note = {arXiv:2106.02643 [astro-ph]},
	pages = {22},
}

@misc{collaboration_gw231123_2025,
	title = {{GW231123}: a {Binary} {Black} {Hole} {Merger} with {Total} {Mass} 190-265 \${M}\_\{{\textbackslash}odot\}\$},
	shorttitle = {{GW231123}},
	url = {http://arxiv.org/abs/2507.08219},
	doi = {10.48550/arXiv.2507.08219},
	urldate = {2025-08-13},
	publisher = {arXiv},
	author = {Abac et al., A.G.},
	month = aug,
	year = {2025},
	note = {arXiv:2507.08219 [astro-ph]},
}

@misc{borchers_gravitational-wave_2025,
	title = {Gravitational-wave kicks impact spins of black holes from hierarchical mergers},
	url = {http://arxiv.org/abs/2503.21278},
	doi = {10.48550/arXiv.2503.21278},
	urldate = {2025-08-13},
	publisher = {arXiv},
	author = {Borchers, Angela and Ye, Claire S. and Fishbach, Maya},
	month = mar,
	year = {2025},
	note = {arXiv:2503.21278 [astro-ph]},
}

@article{ligoscientificcollaborationGWTC3CompactBinary2023,
	title = {{GWTC}-3: {Compact} {Binary} {Coalescences} {Observed} by {LIGO} and {Virgo} during the {Second} {Part} of the {Third} {Observing} {Run}},
	volume = {13},
	shorttitle = {{GWTC}-3},
	url = {https://link.aps.org/doi/10.1103/PhysRevX.13.041039},
	doi = {10.1103/PhysRevX.13.041039},
	number = {4},
	urldate = {2024-01-22},
	journal = {Phys. Rev. X},
	author = {Abbott et al., R.},
	month = dec,
	year = {2023},
	note = {Publisher: American Physical Society},
	pages = {041039},
}

@article{thompson_phenomxo4a_2024,
	title = {{PhenomXO4a}: a phenomenological gravitational-wave model for precessing black-hole binaries with higher multipoles and asymmetries},
	volume = {109},
	issn = {2470-0010, 2470-0029},
	shorttitle = {{PhenomXO4a}},
	url = {http://arxiv.org/abs/2312.10025},
	doi = {10.1103/PhysRevD.109.063012},
	number = {6},
	urldate = {2025-08-20},
	journal = {Physical Review D},
	author = {Thompson, Jonathan E. and Hamilton, Eleanor and London, Lionel and Ghosh, Shrobana and Kolitsidou, Panagiota and Hoy, Charlie and Hannam, Mark},
	month = mar,
	year = {2024},
	note = {arXiv:2312.10025 [gr-qc]},
	pages = {063012},
}

@article{varma_surrogate_2019,
	title = {Surrogate models for precessing binary black hole simulations with unequal masses},
	volume = {1},
	issn = {2643-1564},
	url = {http://arxiv.org/abs/1905.09300},
	doi = {10.1103/PhysRevResearch.1.033015},
	number = {3},
	urldate = {2025-08-20},
	journal = {Physical Review Research},
	author = {Varma, Vijay and Field, Scott E. and Scheel, Mark A. and Blackman, Jonathan and Gerosa, Davide and Stein, Leo C. and Kidder, Lawrence E. and Pfeiffer, Harald P.},
	month = oct,
	year = {2019},
	note = {arXiv:1905.09300 [gr-qc]},
	pages = {033015},
}

@article{gerosa_hierarchical_2021,
	title = {Hierarchical mergers of stellar-mass black holes and their gravitational-wave signatures},
	volume = {5},
	copyright = {2021 Springer Nature Limited},
	issn = {2397-3366},
	url = {https://www.nature.com/articles/s41550-021-01398-w},
	doi = {10.1038/s41550-021-01398-w},
	language = {en},
	number = {8},
	urldate = {2025-08-21},
	journal = {Nature Astronomy},
	author = {Gerosa, Davide and Fishbach, Maya},
	month = aug,
	year = {2021},
	note = {Publisher: Nature Publishing Group},
	pages = {749--760},
}

@article{tichy_final_2008,
	title = {Final mass and spin of black-hole mergers},
	volume = {78},
	url = {https://link.aps.org/doi/10.1103/PhysRevD.78.081501},
	doi = {10.1103/PhysRevD.78.081501},
	number = {8},
	urldate = {2025-08-21},
	journal = {Physical Review D},
	author = {Tichy, Wolfgang and Marronetti, Pedro},
	month = oct,
	year = {2008},
	note = {Publisher: American Physical Society},
	pages = {081501},
}

@article{hoy_accelerating_2022,
	title = {Accelerating multimodel {Bayesian} inference, model selection, and systematic studies for gravitational wave astronomy},
	volume = {106},
	issn = {2470-0010, 2470-0029},
	url = {https://link.aps.org/doi/10.1103/PhysRevD.106.083003},
	doi = {10.1103/PhysRevD.106.083003},
	language = {en},
	number = {8},
	urldate = {2025-08-21},
	journal = {Physical Review D},
	author = {Hoy, Charlie},
	month = oct,
	year = {2022},
	pages = {083003},
}

@article{hoy_incorporation_2025,
	title = {Incorporation of model accuracy in gravitational wave {Bayesian} inference},
	volume = {9},
	issn = {2397-3366},
	url = {http://arxiv.org/abs/2409.19404},
	doi = {10.1038/s41550-025-02579-7},
	number = {8},
	urldate = {2025-08-21},
	journal = {Nature Astronomy},
	author = {Hoy, Charlie and Akcay, Sarp and Uilliam, Jake Mac and Thompson, Jonathan E.},
	month = jul,
	year = {2025},
	note = {arXiv:2409.19404 [gr-qc]},
	pages = {1256--1267},
}

@article{cornishBayeswaveBayesianInference2015,
	title = {Bayeswave: {Bayesian} inference for gravitational wave bursts and instrument glitches},
	volume = {32},
	issn = {0264-9381},
	shorttitle = {Bayeswave},
	url = {https://dx.doi.org/10.1088/0264-9381/32/13/135012},
	doi = {10.1088/0264-9381/32/13/135012},
	language = {en},
	number = {13},
	urldate = {2024-10-15},
	journal = {Class. Quantum Grav.},
	author = {Cornish, Neil J. and Littenberg, Tyson B.},
	month = jun,
	year = {2015},
	note = {Publisher: IOP Publishing},
	pages = {135012},
}

@article{cornish_towards_2013,
	title = {Towards a unified treatment of gravitational-wave data analysis},
	volume = {87},
	copyright = {http://link.aps.org/licenses/aps-default-license},
	issn = {1550-7998, 1550-2368},
	url = {https://link.aps.org/doi/10.1103/PhysRevD.87.122003},
	doi = {10.1103/PhysRevD.87.122003},
	language = {en},
	number = {12},
	urldate = {2025-08-22},
	journal = {Physical Review D},
	author = {Cornish, Neil J. and Romano, Joseph D.},
	month = jun,
	year = {2013},
	pages = {122003},
}

@article{thrane_introduction_2019,
	title = {An introduction to {Bayesian} inference in gravitational-wave astronomy: parameter estimation, model selection, and hierarchical models},
	volume = {36},
	issn = {1323-3580, 1448-6083},
	shorttitle = {An introduction to {Bayesian} inference in gravitational-wave astronomy},
	url = {http://arxiv.org/abs/1809.02293},
	doi = {10.1017/pasa.2019.2},
	urldate = {2025-08-22},
	journal = {Publications of the Astronomical Society of Australia},
	author = {Thrane, Eric and Talbot, Colm},
	year = {2019},
	note = {arXiv:1809.02293 [astro-ph]},
	pages = {e010},
}

@article{whittle_analysis_1953,
	title = {The {Analysis} of {Multiple} {Stationary} {Time} {Series}},
	volume = {15},
	issn = {0035-9246},
	url = {https://www.jstor.org/stable/2983728},
	number = {1},
	urldate = {2025-08-22},
	journal = {Journal of the Royal Statistical Society. Series B (Methodological)},
	author = {Whittle, P.},
	year = {1953},
	note = {Publisher: [Royal Statistical Society, Oxford University Press]},
	pages = {125--139},
}

@article{collaboration_gwtc-1_2019,
	title = {{GWTC}-1: {A} {Gravitational}-{Wave} {Transient} {Catalog} of {Compact} {Binary} {Mergers} {Observed} by {LIGO} and {Virgo} during the {First} and {Second} {Observing} {Runs}},
	volume = {9},
	issn = {2160-3308},
	shorttitle = {{GWTC}-1},
	url = {http://arxiv.org/abs/1811.12907},
	doi = {10.1103/PhysRevX.9.031040},
	number = {3},
	urldate = {2025-08-21},
	journal = {Physical Review X},
	author = {Abbott et al., B. P.},
	month = sep,
	year = {2019},
	note = {arXiv:1811.12907 [astro-ph]},
	pages = {031040},
}

@article{abbott_gwtc-2_2021,
	title = {{GWTC}-2: {Compact} {Binary} {Coalescences} {Observed} by {LIGO} and {Virgo} {During} the {First} {Half} of the {Third} {Observing} {Run}},
	volume = {11},
	issn = {2160-3308},
	shorttitle = {{GWTC}-2},
	url = {http://arxiv.org/abs/2010.14527},
	doi = {10.1103/PhysRevX.11.021053},
	number = {2},
	urldate = {2025-08-21},
	journal = {Physical Review X},
	author = {Abbott et al., R.},
	month = jun,
	year = {2021},
	note = {arXiv:2010.14527 [gr-qc]},
	pages = {021053},
}

@article{ashtonBilbyUserfriendlyBayesian2019,
	title = {Bilby: {A} {User}-friendly {Bayesian} {Inference} {Library} for {Gravitational}-wave {Astronomy}},
	volume = {241},
	issn = {0067-0049},
	shorttitle = {Bilby},
	url = {https://dx.doi.org/10.3847/1538-4365/ab06fc},
	doi = {10.3847/1538-4365/ab06fc},
	language = {en},
	number = {2},
	urldate = {2023-10-09},
	journal = {ApJS},
	author = {Ashton, Gregory and Hübner, Moritz and Lasky, Paul D. and Talbot, Colm and Ackley, Kendall and Biscoveanu, Sylvia and Chu, Qi and Divakarla, Atul and Easter, Paul J. and Goncharov, Boris and Vivanco, Francisco Hernandez and Harms, Jan and Lower, Marcus E. and Meadors, Grant D. and Melchor, Denyz and Payne, Ethan and Pitkin, Matthew D. and Powell, Jade and Sarin, Nikhil and Smith, Rory J. E. and Thrane, Eric},
	month = apr,
	year = {2019},
	note = {Publisher: The American Astronomical Society},
	pages = {27},
}

@article{speagleDynestyDynamicNested2020a,
	title = {dynesty: {A} {Dynamic} {Nested} {Sampling} {Package} for {Estimating} {Bayesian} {Posteriors} and {Evidences}},
	volume = {493},
	issn = {0035-8711, 1365-2966},
	shorttitle = {dynesty},
	url = {http://arxiv.org/abs/1904.02180},
	doi = {10.1093/mnras/staa278},
	number = {3},
	urldate = {2023-11-02},
	journal = {Monthly Notices of the Royal Astronomical Society},
	author = {Speagle, Joshua S.},
	month = apr,
	year = {2020},
	note = {arXiv:1904.02180 [astro-ph, stat]},
	pages = {3132--3158},
}

@article{mapelli_hierarchical_2021,
	title = {Hierarchical black hole mergers in young, globular and nuclear star clusters: the effect of metallicity, spin and cluster properties},
	volume = {505},
	issn = {0035-8711},
	shorttitle = {Hierarchical black hole mergers in young, globular and nuclear star clusters},
	url = {https://doi.org/10.1093/mnras/stab1334},
	doi = {10.1093/mnras/stab1334},
	number = {1},
	urldate = {2025-08-22},
	journal = {Monthly Notices of the Royal Astronomical Society},
	author = {Mapelli, Michela and Dall’Amico, Marco and Bouffanais, Yann and Giacobbo, Nicola and Arca Sedda, Manuel and Artale, M Celeste and Ballone, Alessandro and Di Carlo, Ugo N and Iorio, Giuliano and Santoliquido, Filippo and Torniamenti, Stefano},
	month = jul,
	year = {2021},
	pages = {339--358},
}

@article{kimball_evidence_2021,
	title = {Evidence for {Hierarchical} {Black} {Hole} {Mergers} in the {Second} {LIGO}–{Virgo} {Gravitational} {Wave} {Catalog}},
	volume = {915},
	issn = {2041-8205},
	url = {https://dx.doi.org/10.3847/2041-8213/ac0aef},
	doi = {10.3847/2041-8213/ac0aef},
	language = {en},
	number = {2},
	urldate = {2025-08-22},
	journal = {The Astrophysical Journal Letters},
	author = {Kimball, Chase and Talbot, Colm and Berry, Christopher P L and Zevin, Michael and Thrane, Eric and Kalogera, Vicky and Buscicchio, Riccardo and Carney, Matthew and Dent, Thomas and Middleton, Hannah and Payne, Ethan and Veitch, John and Williams, Daniel},
	month = jul,
	year = {2021},
	note = {Publisher: The American Astronomical Society},
	pages = {L35},
}

@article{fragione_origin_2020,
	title = {On the {Origin} of {GW190521}-like {Events} from {Repeated} {Black} {Hole} {Mergers} in {Star} {Clusters}},
	volume = {902},
	issn = {2041-8205},
	url = {https://dx.doi.org/10.3847/2041-8213/abbc0a},
	doi = {10.3847/2041-8213/abbc0a},
	language = {en},
	number = {1},
	urldate = {2025-08-22},
	journal = {The Astrophysical Journal Letters},
	author = {Fragione, Giacomo and Loeb, Abraham and Rasio, Frederic A.},
	month = oct,
	year = {2020},
	note = {Publisher: The American Astronomical Society},
	pages = {L26},
}

@article{rodriguez_black_2019,
	title = {Black holes: {The} next generation---repeated mergers in dense star clusters and their gravitational-wave properties},
	volume = {100},
	shorttitle = {Black holes},
	url = {https://link.aps.org/doi/10.1103/PhysRevD.100.043027},
	doi = {10.1103/PhysRevD.100.043027},
	number = {4},
	urldate = {2025-08-22},
	journal = {Physical Review D},
	author = {Rodriguez, Carl L. and Zevin, Michael and Amaro-Seoane, Pau and Chatterjee, Sourav and Kremer, Kyle and Rasio, Frederic A. and Ye, Claire S.},
	month = aug,
	year = {2019},
	note = {Publisher: American Physical Society},
	pages = {043027},
}

@article{liu_hierarchical_2021,
	title = {Hierarchical black hole mergers in multiple systems: constrain the formation of {GW190412}-, {GW190814}-, and {GW190521}-like events},
	volume = {502},
	issn = {0035-8711},
	shorttitle = {Hierarchical black hole mergers in multiple systems},
	url = {https://doi.org/10.1093/mnras/stab178},
	doi = {10.1093/mnras/stab178},
	number = {2},
	urldate = {2025-08-22},
	journal = {Monthly Notices of the Royal Astronomical Society},
	author = {Liu, Bin and Lai, Dong},
	month = apr,
	year = {2021},
	pages = {2049--2064},
}

@article{dallamico_gw190521_2021,
	title = {{GW190521} formation via three-body encounters in young massive star clusters},
	volume = {508},
	issn = {0035-8711},
	url = {https://doi.org/10.1093/mnras/stab2783},
	doi = {10.1093/mnras/stab2783},
	number = {2},
	urldate = {2025-08-22},
	journal = {Monthly Notices of the Royal Astronomical Society},
	author = {Dall’Amico, Marco and Mapelli, Michela and Di Carlo, Ugo N and Bouffanais, Yann and Rastello, Sara and Santoliquido, Filippo and Ballone, Alessandro and Arca Sedda, Manuel},
	month = dec,
	year = {2021},
	pages = {3045--3054},
}

@article{arca-sedda_breaching_2021,
	title = {Breaching the {Limit}: {Formation} of {GW190521}-like and {IMBH} {Mergers} in {Young} {Massive} {Clusters}},
	volume = {920},
	issn = {0004-637X},
	shorttitle = {Breaching the {Limit}},
	url = {https://dx.doi.org/10.3847/1538-4357/ac1419},
	doi = {10.3847/1538-4357/ac1419},
	language = {en},
	number = {2},
	urldate = {2025-08-22},
	journal = {The Astrophysical Journal},
	author = {Arca-Sedda, Manuel and Paolo Rizzuto, Francesco and Naab, Thorsten and Ostriker, Jeremiah and Giersz, Mirek and Spurzem, Rainer},
	month = oct,
	year = {2021},
	note = {Publisher: The American Astronomical Society},
	pages = {128},
}

@article{torniamenti_hierarchical_2024,
	title = {Hierarchical binary black hole mergers in globular clusters: {Mass} function and evolution with redshift},
	volume = {688},
	copyright = {https://creativecommons.org/licenses/by/4.0},
	issn = {0004-6361, 1432-0746},
	shorttitle = {Hierarchical binary black hole mergers in globular clusters},
	url = {https://www.aanda.org/10.1051/0004-6361/202449272},
	doi = {10.1051/0004-6361/202449272},
	language = {en},
	urldate = {2025-08-22},
	journal = {Astronomy \& Astrophysics},
	author = {Torniamenti, Stefano and Mapelli, Michela and Périgois, Carole and Arca Sedda, Manuel and Artale, Maria Celeste and Dall’Amico, Marco and Vaccaro, Maria Paola},
	month = aug,
	year = {2024},
	pages = {A148},
}

@article{arcasedda_isolated_2023,
	title = {Isolated and dynamical black hole mergers with {B}-{POP} : the role of star formation and dynamics, star cluster evolution, natal kicks, mass and spins, and hierarchical mergers},
	volume = {520},
	copyright = {https://academic.oup.com/journals/pages/open\_access/funder\_policies/chorus/standard\_publication\_model},
	issn = {0035-8711, 1365-2966},
	shorttitle = {Isolated and dynamical black hole mergers with {B}-{POP}},
	url = {https://academic.oup.com/mnras/article/520/4/5259/7009215},
	doi = {10.1093/mnras/stad331},
	language = {en},
	number = {4},
	urldate = {2025-08-22},
	journal = {Monthly Notices of the Royal Astronomical Society},
	author = {Arca Sedda, Manuel and Mapelli, Michela and Benacquista, Matthew and Spera, Mario},
	month = feb,
	year = {2023},
	pages = {5259--5282},
}

@article{qin_spin_2018,
	title = {The spin of the second-born black hole in coalescing binary black holes},
	volume = {616},
	issn = {0004-6361},
	url = {https://ui.adsabs.harvard.edu/abs/2018A&A...616A..28Q},
	doi = {10.1051/0004-6361/201832839},
	urldate = {2025-08-29},
	journal = {Astronomy and Astrophysics},
	author = {Qin, Y. and Fragos, T. and Meynet, G. and Andrews, J. and Sørensen, M. and Song, H. F.},
	month = aug,
	year = {2018},
	note = {ADS Bibcode: 2018A\&A...616A..28Q},
	pages = {A28},
}


\end{document}